\documentclass[aps,prl,twocolumn,showpacs,superscriptaddress]{revtex4}

\usepackage{amssymb}
\usepackage{amsmath}
\usepackage{mathrsfs}
\usepackage[british]{babel}
\usepackage{color}
\usepackage{graphicx}

\begin{document}

\title{Searching with and against the stream: L{\'e}vy or Brown?}

\author{Vladimir V. Palyulin}
\affiliation{Institute for Physics \& Astronomy, University of Potsdam,
D-14476 Potsdam-Golm, Germany}
\author{Aleksei V. Chechkin}
\affiliation{Akhiezer Institute for Theoretical Physics NSC KIPT,
Kharkov 61108, Ukraine}
\affiliation{Max Planck Institute for the Physics of Complex Systems, D-01187
Dresden, Germany}
\author{Ralf Metzler}
\affiliation{Institute for Physics \& Astronomy, University of Potsdam,
D-14476 Potsdam-Golm, Germany}
\affiliation{Physics Department, Tampere University of Technology,
FI-33101 Tampere, Finland}

\date{\today}

\begin{abstract}
We study the efficiency of search processes based on L{\'e}vy flights (LFs)
with power-law distributed jump lengths in the presence of an external drift.
While LFs turn out to be efficient search processes when relative to the
starting point the target is upstream, in the downstream scenario regular
Brownian motion turns out to be advantageous. This is caused by the
occurrence of leapovers of LFs, due to which LFs typically overshoot a point
in space. We establish criteria when the combination of
the external stream and the initial distance between the starting point and
the target favors LFs over regular Brownian search. Contrary to the common
belief that LFs with a stable index $\alpha=1$ are optimal, we find that the
optimal $\alpha$ may range in the entire interval $(1,2)$ and even include
Brownian motion as the overall most efficient search strategy.
\end{abstract}


\maketitle

How do you find a needle in a haystack? Scientists have studied the
dynamics and optimization of search processes for decades, their interest
ranging from military tasks such as locating enemy submarines, over search
strategies of animals for food, to diffusion control of molecular processes
in biological
cells \cite{olivier,gandhi}. Without prior knowledge about the location of
the target, a searcher randomly explores the search space. However, as
already argued by Shlesinger and Klafter \cite{shlekla}, instead of performing
a Brownian walk a better search strategy for sparse targets is that of a
L{\'e}vy flight (LF): the agent moves randomly with a power-law distribution
$\lambda(x)\simeq|x|^{-1-\alpha}$ of relocation lengths. Due to their lack of
a length scale, LFs combine local exploration with decorrelating, long-range
excursions, and are thus more efficient than searchers following a Gaussian
form of $\lambda(x)$ \cite{viswanathan,bartumeus}.

There exist competing random search models, such as intermittent dynamics
switching between local diffusive search and ballistic relocations
\cite{intermittent}, or persistent random walk models \cite{persist}.
However, while the difference in performance is small \cite{michael}, the
central advantage of LF strategies is the robustness: while other models work
best when their parameters are optimized for specific environmental conditions
(e.g., the target density), LFs remain close to optimal even when these
conditions change \cite{michael}. LFs are thus a preferred
strategy when there is insufficient prior knowledge on the
search space. Indeed, power-law relocation statistics were observed for a
variety of species, including mussels \cite{mussels}, plant lice \cite{aphids},
bats \cite{bats}, marine predators \cite{sharks}, spider monkeys
\cite{spider}, and even for human motion patterns \cite{humans}.
LFs also emerge naturally in models for molecular gene regulation
\cite{michael1}. We note that LFs in the biological context are often
categorized as saltatory motion \cite{james}.

What happens when the search process is biased? This may occur naturally,
when sharks search in areas with an underwater stream or bats forage on a windy
night. Similarly, this may happen in search algorithms when the complex search
space has an overall tilt. As we show here based on a new definition
of the search efficiency relevant for a single target, the answer to the
question for the optimal search strategy crucially
depends on the presence of such streams, in particular, whether the stream is
towards or away from the target. We also show that complementary criteria
for the optimization of the search process
lead to different answers for what is the best search strategy. Thus, Brownian
search may be more efficient
than LF search when the stream is towards the target or, alternatively, when
the target happens to be close to the searcher. Conversely, LF search wins out
when the target is difficult to locate. Our results shed new light on the
long-standing question of optimization in random search processes.

\begin{figure}
\begin{center}
\includegraphics[width=8cm]{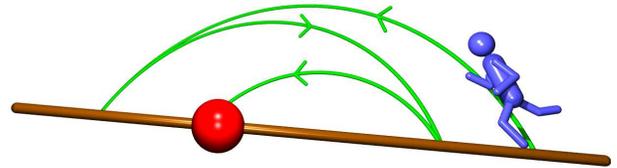}
\end{center}
\caption{Scheme of the search process. A random walker performs random jumps in
the search space until hitting the target. Here, the search is biased by a drift
away from the target. Such an uphill drift caused, for instance, by underwater
streams or above-ground winds affects the search efficiency.}
\label{scheme}
\end{figure}

\textit{First arrival time}. Consider the scenario sketched in Fig.~\ref{scheme}.
A random walker searches for the target by performing random jumps. These are
biased by an external drift. We refer to the bias as \emph{downhill\/} when the
target lies in the direction of the stream as seen from the initial position of
the random walker and vice versa. Discovery of the target then corresponds
to the process of first arrival of the walker at the target position. We recall
that for LFs long leapovers with length distribution $p(\ell)\simeq\ell^{-1-
\alpha/2}$ across a point may frequently occur, and thus the
probability to actually arrive at a point is significantly smaller than the
passage of the walker across this point \cite{koren,fpt}. As basis for our
description we use the fractional Fokker-Planck equation (FFPE) for LFs in
the presence of the drift velocity $v$ \cite{report},
\begin{equation}
\frac{\partial f(x,t)}{\partial t}=\frac{\partial^{\alpha}f(x,t)}{
\partial \left\vert x \right\vert^{\alpha}}-v\frac{\partial f(x,t)}{\partial
x}-\wp_{\mathrm{fa}}(t)\delta(x)
\label{ffpe}
\end{equation}
defined for $0<\alpha\le2$. The distribution $f(x,t)$ is the density function to
find the walker at position $x$ at time $t$, for which we assume the initial
position $x_0$, i.e., $f(x,0)=\delta(x-x_0)$. The fractional derivative
$\partial^{\alpha}/\partial|x|^{\alpha}$ is defined in terms of its Fourier
transform, $\int_{-\infty}^{\infty}e^{ikx}\partial^{\alpha}/\partial|x|^{\alpha}
f(x,t)dx=-|k|^{\alpha}f(k,t)$, where $f(k,t)=\int_{-\infty}^{\infty}e^{ikx}f(x,t)
dx$ is the Fourier transform of $f(x,t)$ \cite{REM}. Thus, in the limit $\alpha
=2$ we recover the standard Fokker-Planck equation of Brownian motion. In
Eq.~(\ref{ffpe}), we
introduced rescaled, dimensionless variables, such that $v$ is a measure for the
amplitude of the drift (see the Supplementary Material for the rescaling of the
FFPE \cite{supp}). In Eq.~(\ref{ffpe}) we implemented a point sink at $x=0$
representing the target: the random walker is removed when the target is hit.
Here, $\wp_{\mathrm{fa}}(t)$ is the density of first
arrival. Eq.~(\ref{ffpe}) generalizes the first arrival dynamics in absence of
a drift of Ref.~\cite{fpt}.
Due to the sink term, the density function $f(x,t)$ is not normalized, that is,
the cumulative survival $\mathscr{S}(t)=\int_{-\infty}^{\infty}f(x,t)dx$ is a
decreasing function of time. Using the properties of the fractional derivative,
integration of Eq.~(\ref{ffpe}) over the position coordinate $x$ delivers the
first arrival density, $\wp_{\mathrm{fa}}(t)=-(d/dt)\int_{-\infty
}^{\infty}f(x,t)dx=-(d/dt)\mathscr{S}(t)$.

The solution of Eq.~(\ref{ffpe}) can be obtained via Fourier-Laplace transform,
and for $\wp_{\mathrm{fa}}$ we find
\begin{equation}
\wp_{\mathrm{fa}}(s)=\int_{-\infty}^{\infty}e^{ikx_0}\xi dk\Big/
\int_{-\infty}^{\infty}\xi dk,
\xi=(s+|k|^{\alpha}-ikv)^{-1},
\label{pfa}
\end{equation}
where the Laplace transform is defined as $f(x,s)=\int_0^{\infty}e^{-st}f(x,t)
dt$. Result (\ref{pfa}) instantly shows an important feature: for discontinuous
LFs with $0<\alpha\le1$, the quantity $\wp_{\mathrm{fa}}(s)$ vanishes, since
the integral in the denominator diverges while the integral in the numerator
converges. Thus L{\'e}vy search for a point-like target will never succeed for
$0<\alpha\le1$. This property reflects transience of L{\'e}vy flights with
$\alpha<d$, where $d$ is the embedding spatial dimension \cite{sato}.
In this sense the value
$\alpha=1$ obtained for optimal search for sparse targets in drift-free search
\cite{gandhi,michael1} is to be seen as limiting point of $\alpha$
from above unity. We obtained analytical results for the first arrival behavior
encoded in Eq.~(\ref{pfa}) in the limit of a small bias, see SM \cite{supp}. In
the following we combine numerical analysis and complementary definitions of the
search efficiency to study the optimal random search of Brownian versus LF
strategies.

\textit{Search efficiency.} What is a good measure for the efficiency of a search
mechanism? There are two frequently used definitions of search
efficiency, counting the number of found targets either per traveled unit
distance or per number of steps \cite{james}. These definitions work well when
there is a finite target density. Here we are interested in the more natural
problem of search for a single target, a countable number of targets, or a
finite target area. In such cases the average search time diverges, and we thus
need a modified definition for the search efficiency. We choose the average over
inverse search times,
\begin{equation}
\mathcal{E}=\left<\frac{1}{t}\right>=\int_0^{\infty}\wp_{\mathrm{fa}}(s)ds,
\label{eff}
\end{equation}
where $\langle\cdot\rangle=\int_0^{\infty}\cdot\wp_{\mathrm{fa}}(t)dt$. Due to
the definition of $\mathcal{E}$ as inverse first arrival times, contributions
from short and intermediate times dominate the efficiency.
To demonstrate the usefulness of definition (\ref{eff}) we
determined $\mathcal{E}$ for a Brownian searcher for both downhill and uphill
situations with arbitrary $v$ and $x_0>0$. We find respectively,
\begin{equation}
\label{EffBrow}
\mathcal{E}=\frac{2}{x_0^2}\left(1+\frac{|v|x_0}{2}\right)\left\{
\begin{array}{ll}1,& v\leq0\\[0.2cm]
\exp\left(-vx_0\right),& v\geq0\end{array}\right.
\end{equation}
Consistently we observe that the search efficiency increases with $v$ when the
stream pushes the searcher towards the target, while the efficiency decreases
exponentially in the uphill case. The latter can be interpreted as an activation
barrier for target detection. In absence of a drift the efficiency is just
the inverse mean diffusion time (on average, $x_0^2\sim2t$ in dimensionless
units).

Combining expressions (S4) and (\ref{eff}) we obtain the search efficiency for
an LF in the presence of a weak bias,
\begin{equation}
\mathcal{E}=\frac{\alpha}{x_0^\alpha}\left[\cos\left(\pi\left[1-\frac{\alpha}{2}
\right]\right)\Gamma(\alpha)-2\left(1-\frac{1}{\alpha}\right)\mathrm{Pe}_{\alpha}
\right],
\label{EffLevyFlat}
\end{equation}
for $1<\alpha\leq2$. Here we introduced the generalized P{\'e}clet number
$\mathrm{Pe}_{\alpha}=vx_0^{\alpha-1}/2$. Note that $\mathrm{Pe}_{\alpha}$ is
in fact dimensionless, due to the rescaling of variables, see Eq.~(\ref{ffpe}).
This is our first main result. In the Brownian limit
$\alpha=2$ the efficiency is $\mathcal{E}\simeq2x_0^{-2}(1-\mathrm{Pe}_2)$,
which corresponds to the small $v$-expansion in Eq.~(\ref{EffBrow}). For
$\alpha\rightarrow1$ and with $x_0$ fixed the efficiency drops to zero.
While $\alpha=1$ is the optimal parameter for LF search of sparse but finite
target density, for the case considered here the transition to
discontinuous LFs at $\alpha=1$ means that the target can no longer be detected.
These observations show that the standard dogmas on the efficiency of random
search processes are much more specific than usually believed.

\begin{figure}
\includegraphics[width=4.2cm]{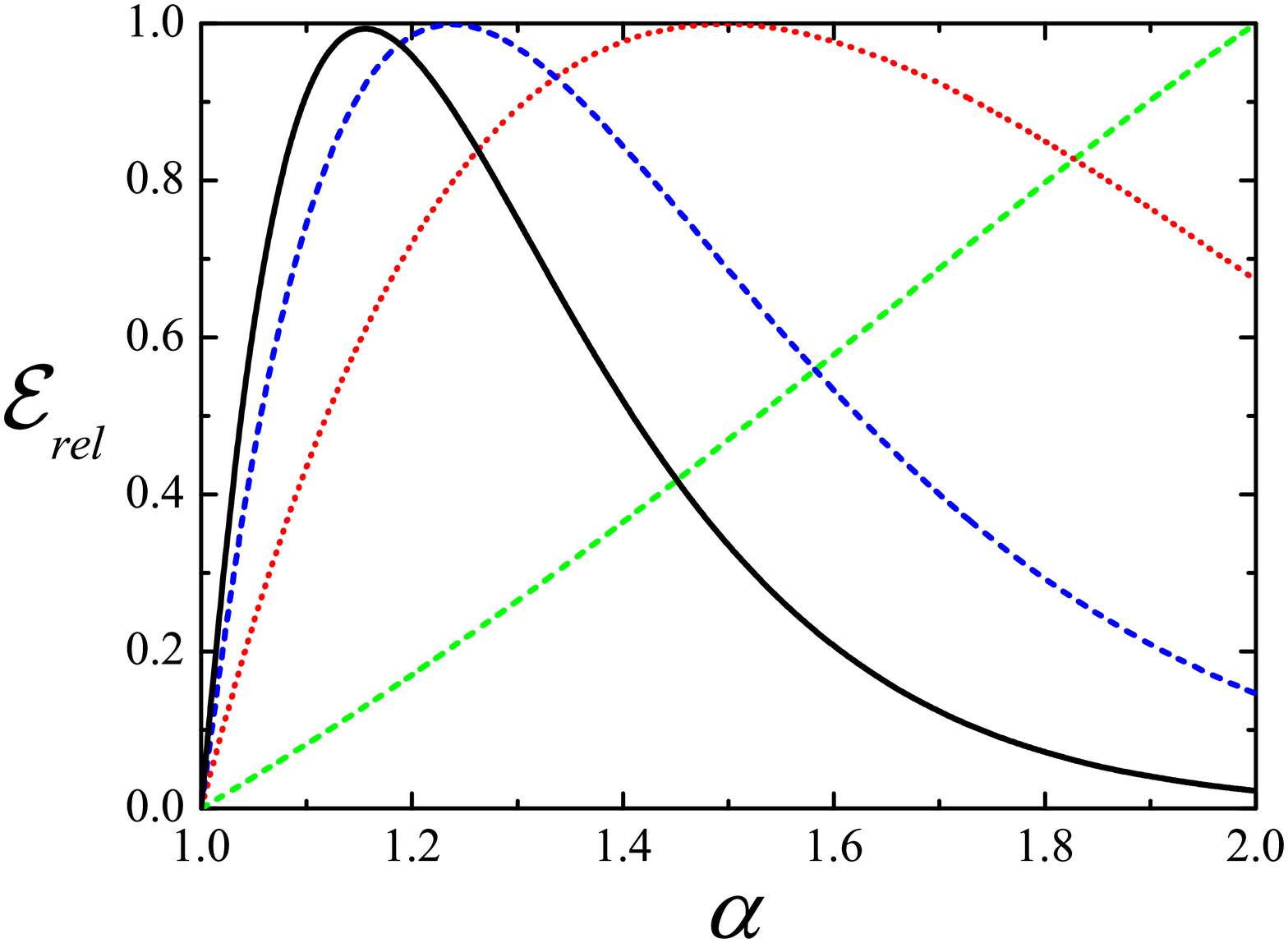}
\includegraphics[width=4.2cm]{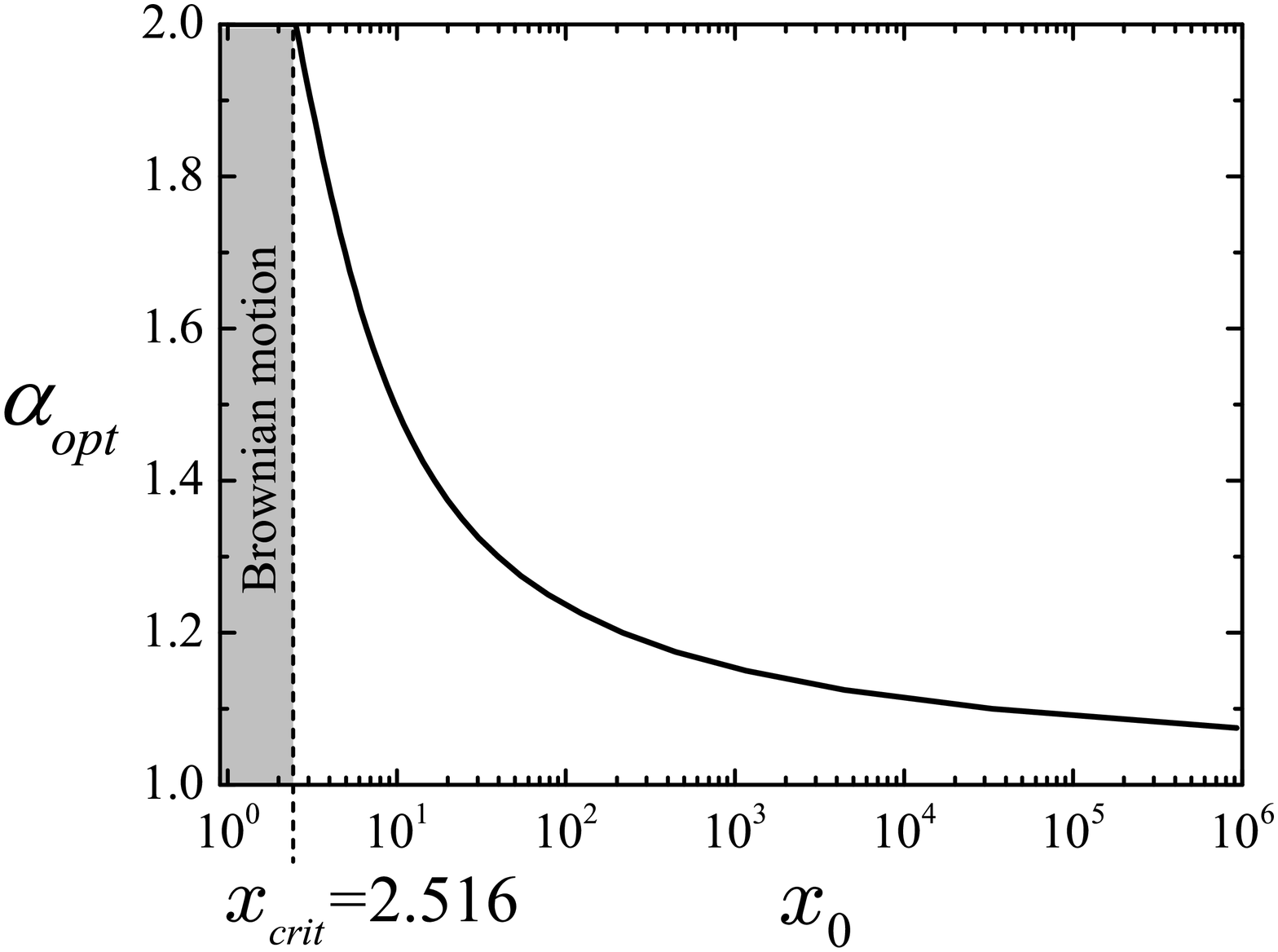}
\caption{Left: Relative efficiency $\mathcal{E}_{\mathrm{rel}}$ for LFs with
$v=0$ as function of stable index $\alpha$, Eq.~(\ref{EffLevyFlat}).
The curves are drawn for the initial positions $x_0=1$ (black
full line), $x_0=10$ (red dashed), $x_0=100$ (blue dotted) and $x_0=1,000$
(pink dash-dotted). With growing $x_0$ the functional shape changes from a
monotonic to non-monotonic shape and then becomes sharper, their
maximum shifting towards unity. Thus, for $x_0=10$ we find $\alpha_{\mathrm{
opt}}\approx1.5$, while for $x_0=1000$, $\alpha_{\mathrm{opt}}\approx1.15$.
Right: Optimal stable index $\alpha_{\mathrm{opt}}$ as function of the
initial position $x_0$, as obtained from Eq.~(\ref{OptAlpha}). For $x_0
\lessapprox2.516$, the optimal search strategy is Brownian (shaded area).}
\label{EffFlatVarAlpha}
\label{EffFlatOptAlpha}
\end{figure}

Let us discuss the efficiency of LF search in more detail, starting with the
case of vanishing drift strength $v$. As the time to reach the target grows
substantially with initial distance $x_0$, we compare the search efficiency at
fixed value $x_0$. In Fig.~\ref{EffFlatVarAlpha} we show the dependence on the
stable index $\alpha$ of the relative efficiency $\mathcal{E}_{\mathrm{rel}}=
\mathcal{E}/\mathcal{E}_{\mathrm{opt}}$, where $\mathcal{E}_{\mathrm{opt}}$ is
the maximal value of $\mathcal{E}$ for given $x_0$ attained at the optimal
stable index $\alpha_{\mathrm{opt}}$. We observe that when the starting point
of the walker $x_0$ is close to the target, the optimal search strategy is
Brownian. This is intuitively clear:
Brownian motion cannot overshoot the target and therefore leads to
quick localization. For more distant targets the oversampling of Brownian
walks, i.e., the tendency to multiply return to previously visited points,
reduces the Brownian efficiency, and LFs win out. This is shown for the larger
$x_0$ values in Fig.~\ref{EffFlatVarAlpha}. Interestingly, the behavior of
$\mathcal{E}_{\mathrm{rel}}$ is non-monotonic, and becomes sharper for
increasing $x_0$. In the limit of very large $x_0$ the optimal value of the
stable index $\alpha$ tends to unity. The non-monotonicity of $\mathcal{E}_{
\mathrm{rel}}$ is one of our central results.

At fixed starting position $x_0$ and in absence of a drift the implicit
expression to determine the
optimal stable index $\alpha_{\mathrm{opt}}$ follows from $\left.d\mathcal{E}/
d\alpha\right|_{\alpha_{\mathrm{opt}}}=0$, the result being
\begin{equation}
x_0=2\exp\left\{\frac{1}{\alpha_{\mathrm{opt}}}+\frac{1}{2}\psi\left(\frac{
\alpha_{\mathrm{opt}}}{2}\right)+\frac{1}{2}\psi\left(\frac{1-\alpha_{\mathrm{
opt}}}{2}\right)\right\}.
\label{OptAlpha}
\end{equation}
Here $\psi$ denotes the digamma function. Eq.~(\ref{OptAlpha}) allows us to
plot $\alpha_{\mathrm{opt}}$ as function of the initial position of the LF
searcher shown in Fig.~\ref{EffFlatOptAlpha}. Interestingly, if for our
dimensionless units the initial position is closer to the target than $x_0
\approx2.516$, then the optimal search strategy is Brownian, otherwise it
corresponds to LFs with $\alpha_{\mathrm{opt}}$.

\begin{figure}
\includegraphics[width=4.2cm]{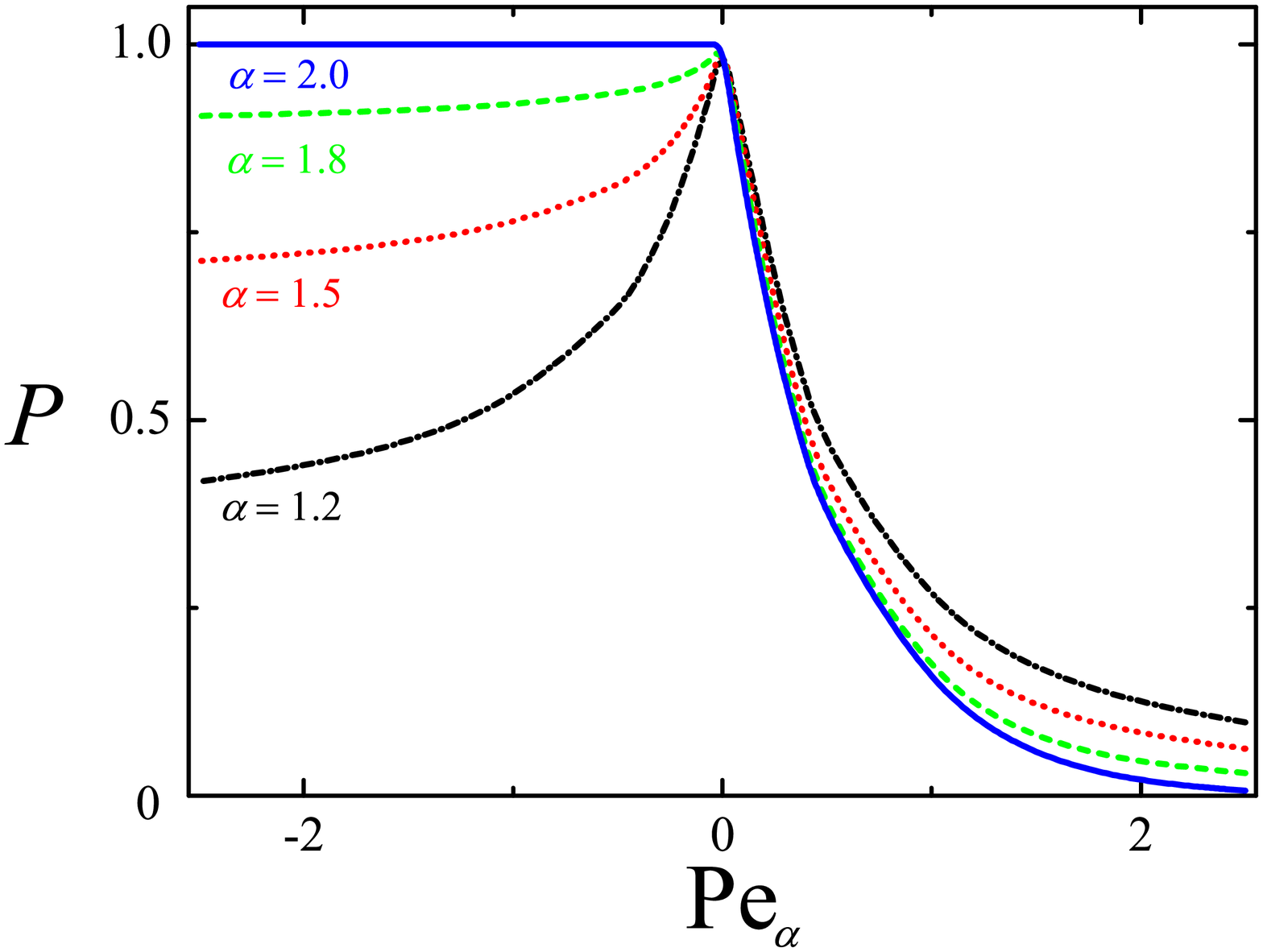}
\includegraphics[width=4.2cm]{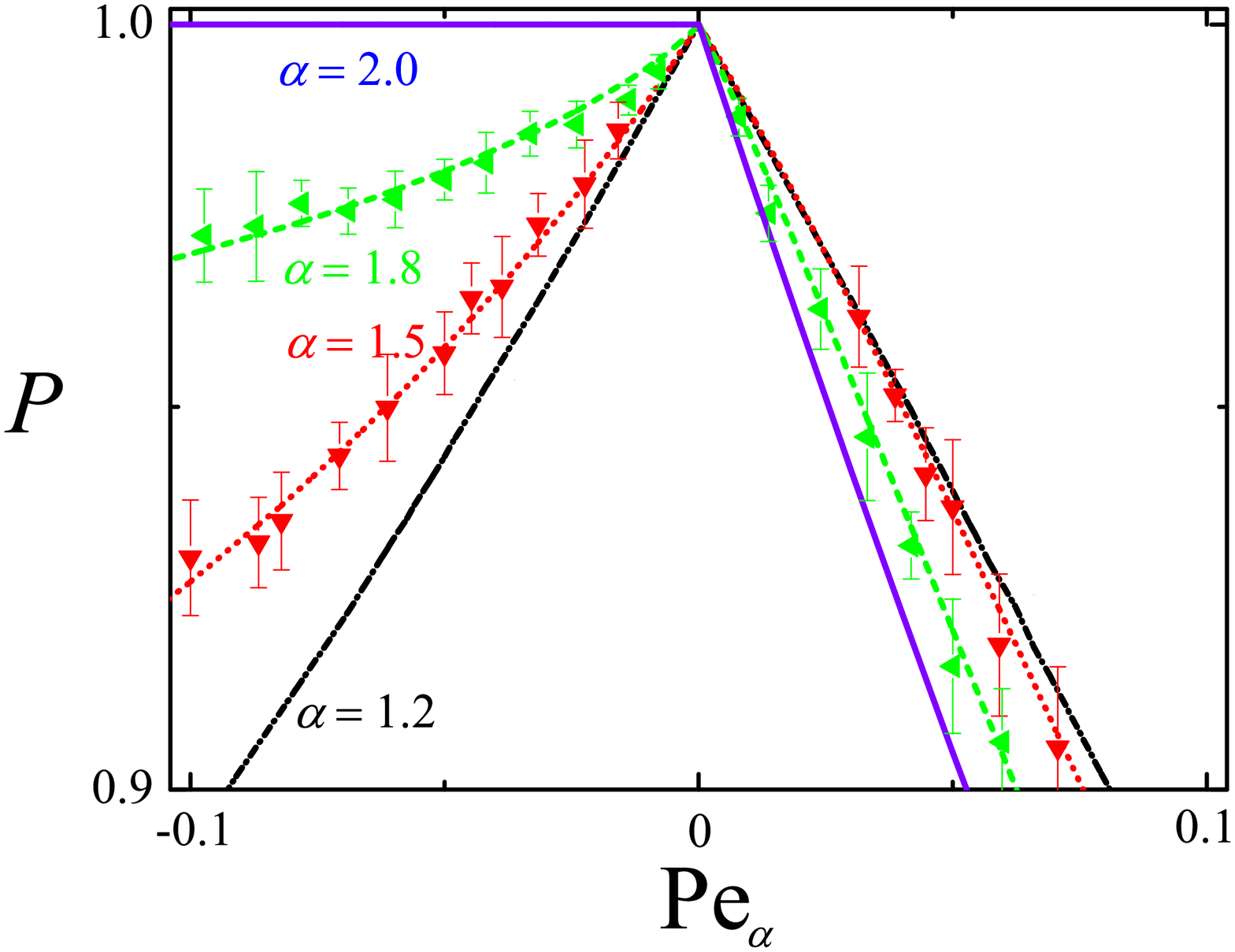}
\caption{Left: Dependence of the cumulative first arrival probability on the
Pecl{\'e}t number $\mathrm{Pe}_{\alpha}$. Lines are from 
numerical solution of Eq.~(\ref{pfa}). The blue (full) line corresponds to
Brownian search, $\alpha=2$, the green (dashed) represents LF search with
$\alpha=1.8$, and the red (dotted) line stands for $\alpha=1.5$. Finally,
the black (dashed-dotted) line is for $\alpha=1.2$. Right: Same for small
values of $\mathrm{Pe}_{\alpha}$. In addition
to the numerical solution of Eq.~(\ref{pfa}), the symbols and error bars
are obtained from Langevin equation simulations of LF trajectories. Coloring
and lineshapes correspond to Fig.~\ref{Pe}.}
\label{Pe}
\label{PeInset}
\end{figure}

Once an external drift is present, the arrival to the target as function of
the initial position $x_0$ and the drift strength $v$ becomes non-trivial. In
particular, there may exist a finite residual survival probability $\lim_{t\to
\infty}\mathscr{S}(t)$. The probability $P=\int_0^{\infty}\wp_{\mathrm{fa}}(t)dt
=1-\lim_{t\to\infty}\mathscr{S}(t)$ to successfully reach the target quantifies
the ability of the process to ever reach the target. For some purposes this
measure may be more relevant than the efficiency $\mathcal{E}$. A large
value of $P$ for given parameters corresponds to a high success probability to
eventually locate the target. $P$
is displayed for a large range of the generalized P{\'e}clet number $\mathrm{Pe}
_{\alpha}$ in Fig.~\ref{Pe}. In addition, Fig.~\ref{PeInset} depicts the
small-$\mathrm{Pe}_{\alpha}$ case. These results
are obtained from numerical solution of Eq.~(\ref{pfa}) and are thus not
restricted to small values of $\mathrm{Pe}_{\alpha}$ \cite{SimulDetails}. From
dimensional analysis it is straightforward to show that the success probability
$P$ solely depends on the single parameter $\mathrm{Pe}_{\alpha}$.

In the downhill case, when the searcher is pushed towards the target by the
external stream ($\mathrm{Pe}_{\alpha}<0$) the best strategy in terms of $P$ is
always that of Brownian search, reaching $P=1$ for all values of $\mathrm{Pe}_{
\alpha}$.
The LF searcher in this regime always fares worse ($P<1$), the discrepancy
increasing for smaller values of the stable index. This is due to the occurrence
of leapovers across the target for LFs. In the presence of a strong drift, the
success probability $P$ becomes considerably smaller.
The opposite tendency is observed for the uphill case when the walker needs
to move against the stream towards the target ($\mathrm{Pe}_{\alpha}>0$). Now,
LFs with a smaller stable index perform better, due to the possibility to
approach the target faster with fewer jumps. We note, however, that the absolute
gain of LF versus Brownian search in the uphill case is considerably smaller
than the loss in the downhill scenario.

\begin{figure}
\includegraphics[width=4.2cm]{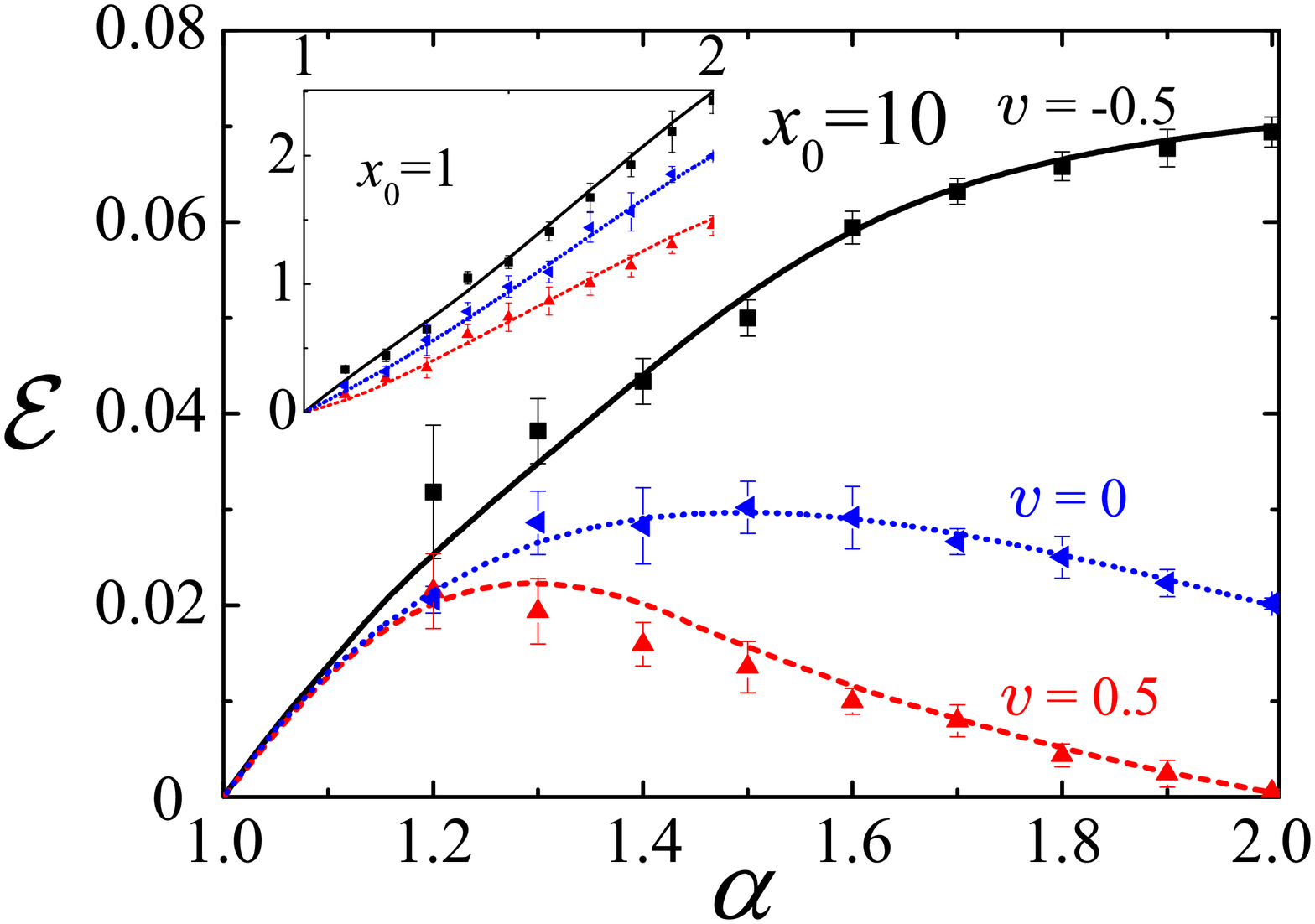}
\includegraphics[width=4.2cm]{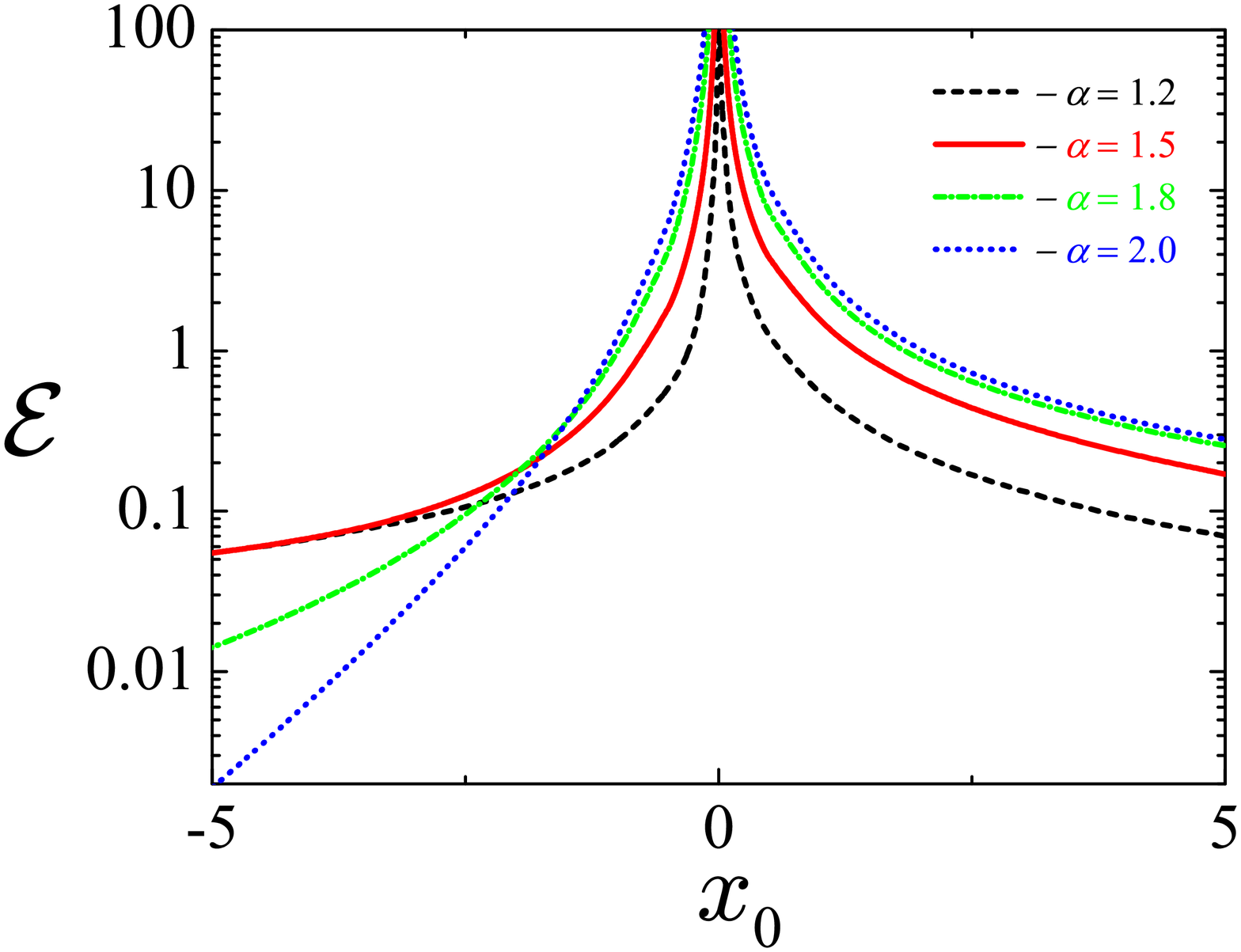}
\caption{Left: Efficiency as a function of stable index $\alpha$ for initial
positions $x_0=10$ and $x=1$ (Inset). We show the downhill case ($v=-0.5$,
full black curve), neutral case ($v=0$, blue dotted curve), and uphill case
($v=0.5$, red dashed curve). Right: Efficiency versus initial
position $x_0$ for $v=-1.0$, i.e., negative $x_0$ correspond to
uphill motion.}
\label{EffLevy}
\end{figure}

The search efficiency $\mathcal{E}$ is affected by the external stream
even more dramatically than the success probability $P$, as shown
in Fig.~\ref{EffLevy}. Here, the initial position is fixed at $x_0=10$ in the
main Figure, and $x_0=1$ in the inset. Black (full) lines correspond to the
downhill case with $v=-0.5$, and the red (dashed) curves to the uphill case
with $v=0.5$. The neutral case $v=0$ is shown by the blue (dotted) line. For
$\alpha\rightarrow1$ the curves converge, in the case $x_0=10$ they almost
coincide below $\alpha\approx1.15$. In the case without bias and $x_0=1$,
consistent with our observations in the drift-free case above, the optimal
strategy remains Brownian (see Fig.~\ref{EffFlatOptAlpha}). In contrast, for
the larger initial separation $x_0=10$ the downhill case the optimal search
strategy is also Brownian, while without bias we found $\alpha_{\mathrm{opt}}
\approx 1.5$. In the uphill case the optimal stable index is shifted to
$\alpha_{\mathrm{opt}}\approx1.3$. The delicate behavior of $\alpha_{\mathrm{
opt}}$ is our other important finding.

\textit{Discussion}. So what is now the best search strategy? As we showed here
this depends crucially on what is more important: to reach the target quickly or
to locate it with the highest likelihood. Moreover, the answer to this question
also depends on the situation, whether there is a single or few targets, or
whether we face a constant density of targets. It will be interesting to
study such questions in L{\'e}vy search models for finite target density.

Specifically, we investigated the performance of LF search models along
or against an external stream. Defining the efficiency as the average inverse
arrival time $\left<1/t\right>$ to the target, we obtained a versatile measure
to quantify search processes when the search space does not have a constant
target density. This efficiency $\left<1/t\right>$ reproduces the features of
Brownian search and works well for both unbiased and biased search processes,
unlike the similar construct $1/\left<t\right>$. In terms of this
efficiency we investigate the optimal search strategy, comparing L{\'e}vy
and Brownian search processes. Without an external bias, it turns out that
the optimal strategy depends on the initial separation between the searcher
and the target: for small separations Brownian motion is the most efficient
way of finding the target. On increasing this separation LFs become more and
more efficient in comparison to Brownian search, and the stable index $\alpha$
decreases towards unity in the limit of very large initial searcher-target
separation. In particular, we find that despite the common claim that LFs with
$\alpha=1$ are most efficient, depending on the parameters of the search space
the optimal stable index may range in the whole interval between unity and two.

When the searcher moves with or against an external stream the analysis in
terms of the success probability $P$ shows that when the initial position of
the searcher with respect to the target is along the stream, the optimal search
strategy is always
Brownian, due to the combined effect of biased motion and absence of the
leapovers in the case of LFs. The average search time is
then simply given in terms of the ratio of initial searcher-target separation 
and the drift velocity $v$. When the searcher needs to reach the target
against the stream, LFs provide the better search strategy. This trend is
confirmed by the results for the success probability $P$. Remarkably,
the gain from using the Brownian strategy instead of LFs in the downhill
scenario is significantly larger than the loss from using Brownian motion
instead of LFs in the uphill case. Depending on the details of the search
space, without prior knowledge on the strength and direction of external
streams, the choice of a Brownian strategy might therefore be overall
advantageous, in contrast to the general dogma in favor of L{\'e}vy search.
These observations may be of particular importance to swimming or airborne
searchers, as streams occur most naturally there. They may also be relevant
for computational search algorithms in biased landscapes.

VVP acknowledges discussions with A. Cherstvy and financial support from
Deutsche Forschungsgemeinschaft. RM acknowledges financial support from the
Academy of Finland (FiDiPro scheme).


\begin{thebibliography}{99}
\bibitem{SimulDetails} Increase of $|\mathrm{Pe}|$ values increases the LE
simulation time. We confirmed FFPE results with LE simulation for $|\mathrm{
Pe}|\leq0.1$ (Fig.~\ref{PeInset}) with very good agreement. Each simulation
point was obtained as a ratio of trapped particles to the overall number of
10,000 runs. We estimated error bars by computation of the deviations $P$ for
each of consecutive 1,000 runs.
\end{thebibliography}
\end{document}